\documentclass[aps,showpacs,twocolumn]{revtex4}
\usepackage{epsfig}

\begin{document}

\title{An alternative approach to the $\sigma$-meson-exchange in nucleon-nucleon interaction}

\author{Lingzhi Chen$^1$,Hourong Pang$^2$, Hongxia Huang$^3$, Jialun Ping$^3$ and Fan Wang$^1$}

\affiliation{$^1$Department of Physics, Nanjing University, Nanjing
210093, P.R. China \\
$^2$Department of Physics, Southeast University, Nanjing 210094,
P.R. China \\
$^3$Department of Physics, Nanjing Normal University, Nanjing
210097, P.R. China}

\begin{abstract}
Through a quantitative comparative study of the properties of
deuteron and nucleon-nucleon interaction with chiral quark model and
quark delocalization color screening model. We show that the
$\sigma$-meson exchange used in the chiral quark model can be
replaced by quark delocalization and color screening mechanism.
\end{abstract}


\maketitle

\section{Introduction}
There has been a long history of studying nucleon-nucleon ($NN$)
interaction since the Yukawa theory. In the effective meson-nucleon
coupling model, after including pseudo-scalar-, vector- and
scalar-meson, especially the $\sigma$-meson, the $NN$ interaction
has been described very well\cite{mah}. In the traditional picture
of $NN$ interaction, the short-range repulsion is provided by
vector-meson ($\rho,\omega$) exchange, the intermediate-range
attraction is accounted for by $\sigma$-meson exchange, and the
long-range tail is attributed to $\pi$ exchange. The pseudo-scalar
and vector mesons are well established, whereas the scalar-meson
$\sigma$ is still in controversy which had been listed, dropped and
relisted in the particle data compilation\cite{pdg}. With the
advancement of Quantum Chromodynamics (QCD), to understand the $NN$
interaction from the fundamental quark-gluon degree of freedom is
expected. However there are only very preliminary lattice QCD
calculations of the $NN$ interaction due to the complexity of low
energy QCD\cite{lat}. Various new approaches have been developed
since the end of 1970's. The chiral perturbation theory described
the $NN$ interaction very well where only $\pi$
($1\pi,2\pi,3\pi,...$) exchange is employed \cite{chPT}. The most
common used quark model in the study of $NN$ interaction is the
chiral quark model\cite{chiral1,chiral2,chiral3}. Both in
meson-nucleon coupling and quark-meson coupling model (chiral quark
model is a typical one) the $\sigma$ meson is indispensable and
which is assumed for a long time to be an effective description of
the correlated $\pi\pi$ resonance. Recently a very broad
$\sigma$-meson as a resonance of $\pi\pi$ was reported\cite{BES}.
However the results, found by three groups independently, show that
the correlated two-pion exchange between two nucleons results in
strong short-range repulsion and very moderate long-range attraction
which is quite different from the intermediate range attraction
resulted from the $\sigma$ meson exchange\cite{sigma}. This raises a
question: Is the $\sigma$-meson used in the one boson exchange model
and chiral quark model the correlated $\pi\pi$ resonance or other
effective one? Is there an alternative approach to account for the
$NN$ intermediate range attraction? In addition, a long standing
fact of the similarity of molecular and nuclear force is well-known,
but no theoretical explanation so far. Is it accidental or is there
a similar mechanism for molecular and nuclear force behind?

in 1990's, A modified version of quark cluster model: quark
delocalization color screening model \cite{QDCSM1}, is proposed. Two
new ingredients were introduced: quark delocalization (to enlarge
the model variational space to take into account the mutual
distortion or the internal excitations of nucleons in the course of
interaction) and color screening (assuming the quark-quark
interaction dependent on quark states aimed to take into account the
QCD effect which has not been included in the two body confinement
and effective one gluon exchange yet). The model has been
successfully applied to describe $NN$, hyperon-nucleon
scattering\cite{QDCSM1}. In this model, the intermediate-range
attraction is achieved by the quark delocalization, which is like
the electron delocalization in molecules. The color screening is
needed to make the quark delocalization effective. The model gave a
natural explanation of the similarity between molecular force and
nuclear force. Since 2000, the model is extended to incorporate the
Goldstone-boson-exchange to describe the long-range tail of the $NN$
interaction\cite{QDCSM2} which is hard to be described with
quark-gluon degree of freedom. Then the only difference between the
chiral quark model\cite{chiral1} and QDCSM is the intermediate-range
attraction mechanism. It is interesting to compare the two
approaches to see whether the effect of $\sigma$-meson can be
replaced by the new mechanism introduced in QDCSM.

After the Introduction, Section II gives a brief description of the
two approaches to $NN$ interaction. The comparison between the two
approaches is presented in Section III. A short summary is given in
the last section.

\section{Two approaches}

\subsection{Chiral quark model}

The Salamanca model was chosen as the representative of the chiral
quark models, because Salamanca group's work covers the hadron
spectra and nucleon-nucleon interaction, and has been extended to
multi-quark states study. The model details can be found in
ref.\cite{Salamanca}. Here only the Hamiltonian and parameters are
given.

The Hamiltonian of Salamanca model in $NN$ sector can be written as
\cite{Salamanca}

\begin{widetext}

\begin{eqnarray}
H &=& \sum_{i=1}^6 \left(m_i+\frac{p_i^2}{2m_i}\right) -T_c
+\sum_{i<j} \left[
V^{OGE}(r_{ij})+V^{\pi}(r_{ij})+V^{\sigma}(r_{ij})+V^{CON}(r_{ij})
\right],
 \nonumber \\
V^{OGE}(r_{ij})&=& \frac{1}{4}\alpha_s \vec{\lambda}_i \cdot
\vec{\lambda}_j
\left[\frac{1}{r_{ij}}-\frac{\pi}{m_q^2}\left(1+\frac{2}{3}\vec{\sigma}_i\cdot\vec{\sigma}_j
\right)
\delta(r_{ij})-\frac{3}{4m_q^2r^3_{ij}}\mathbf{S}_{ij}\right],
\nonumber \\
V^{\pi}(r_{ij})&=& \frac{1}{3}\alpha_{ch}
\frac{\Lambda^2}{\Lambda^2-m_{\pi}^2}m_\pi \left\{ \left[ Y(m_\pi
r_{ij})- \frac{\Lambda^3}{m_{\pi}^3}Y(\Lambda r_{ij}) \right]
\vec{\sigma}_i \cdot\vec{\sigma}_j+\left[ H(m_\pi
r_{ij})-\frac{\Lambda^3}{m_\pi^3}
H(\Lambda r_{ij})\right] \mathbf{S}_{ij} \right\} \vec{\tau}_i \cdot\vec{\tau}_j,  \\
V^{\sigma}(r_{ij})&=& -\alpha_{ch} \frac{4m_q^2}{m_\pi^2}
\frac{\Lambda^2}{\Lambda^2-m_{\pi}^2}m_\sigma \left[ Y(m_\sigma
r_{ij})-\frac{\Lambda}{m_\sigma}Y(\Lambda r_{ij}) \right], ~~~~
 \alpha_{ch}= \frac{g^2_{ch}}{4\pi}\frac{m^2_{\pi}}{4m_u m_u}
 \nonumber \\
V^{CON}(r_{ij})&=& -a_c \vec{\lambda}_i \cdot\vec{\lambda}_j
r^2_{ij}, \nonumber
\\
\mathbf{S}_{ij} & = &  \frac{\vec{\sigma}_i \cdot \vec{r}_{ij}
\vec{\sigma}_j \cdot
\vec{r}_{ij}}{r_{ij}^2}-\frac{1}{3}~\vec{\sigma}_i \cdot
\vec{\sigma}_j. \nonumber
\end{eqnarray}
\end{widetext}
$\mathbf{S}_{ij}$ is quark tensor operator. $Y(x)$ and $H(x)$ are
standard Yukawa functions. $T_c$ is the kinetic energy of the center
of mass. $\alpha_{ch} $ is the chiral coupling constant, which is
determined from $\pi$-nucleon coupling constant as usual. All other
symbols have their usual meanings. The parameters of Hamiltonian are
given in Table I.

\subsection{Quark delocalization, color screening model}

The QDCSM model and its extension were discussed in detail in
ref.\cite{QDCSM1,QDCSM2}. The Hamiltonian and trial wavefunction of
the extended QDCSM are given below.

\begin{widetext}
\begin{eqnarray}
H_6 & = & \sum_{i=1}^6 \left(m_i+\frac{p_i^2}{2m_i}\right)-T_{c}
+\sum_{i<j=1}^{6}
    \left[ V^{CON}(r_{ij}) + V^{OGE}(r_{ij}) +V^{\pi}(r_{ij})
    \right] ,                    \nonumber \\
V^{CON}(r_{ij}) & = & -a_c \vec{\lambda}_i \cdot \vec{\lambda}_j
\left\{ \begin{array}{ll}
 r_{ij}^2 &
 \qquad \mbox{if }i,j\mbox{ occur in the same baryon orbit}, \\
 \frac{1 - e^{-\mu r_{ij}^2} }{\mu} & \qquad
 \mbox{if }i,j\mbox{ occur in different baryon orbits},
 \end{array} \right.
\end{eqnarray}
\end{widetext}
$V^{OGE}$ and $V^{\pi}$ are exact the same as Salamanca's. $\mu$ is
the color screening constant to be determined by fitting the
deuteron mass in this model. The parameters of the model are given
in Table I too.

The quark delocalization in QDCSM is realized by writing the single
particle orbital wave function as a linear combination of left and
right gaussians, the single particle orbital wave functions in the
ordinary quark cluster model.
\begin{eqnarray}
\psi_{\alpha}(\vec{S}_i ,\epsilon) & = & \left(
\phi_{\alpha}(\vec{S}_i)
+ \epsilon \phi_{\alpha}(-\vec{S}_i)\right) /N(\epsilon), \nonumber \\
\psi_{\beta}(-\vec{S}_i ,\epsilon) & = &
\left(\phi_{\beta}(-\vec{S}_i)
+ \epsilon \phi_{\beta}(\vec{S}_i)\right) /N(\epsilon), \nonumber \\
N(\epsilon) & = & \sqrt{1+\epsilon^2+2\epsilon e^{-S_i^2/4b^2}}. \label{1q} \\
\phi_{\alpha}(\vec{S}_i) & = & \left( \frac{1}{\pi b^2}
\right)^{3/4}
   e^{-\frac{1}{2b^2} (\vec{r}_{\alpha} - \vec{S}_i/2)^2} \nonumber \\
\phi_{\beta}(-\vec{S}_i) & = & \left( \frac{1}{\pi b^2}
\right)^{3/4}
   e^{-\frac{1}{2b^2} (\vec{r}_{\beta} + \vec{S}_i/2)^2}. \nonumber
\end{eqnarray}

\begin{center}
Table I. The parameters of two models.

\begin{tabular}{ccccc}\hline\hline
 & Salamanca Model & \multicolumn{3}{c}{QDCSM}  \\
 & & set 1 & set 2 & set 3 \\ \hline
$m_{u,d}(MeV)$ & 313 & & 313 & \\
$\alpha_{ch}$ & 0.027 & & 0.027 & \\
$m_\pi(MeV)$ & 138 & & 138 & \\
$\Lambda (fm^{-1})$ & 4.2 & & 4.2 & \\ \hline
$\alpha_s$ & 0.485 & 0.485 & 0.724 & 0.996 \\
$b (fm)$ & 0.518 & 0.518 & 0.56 & 0.60 \\
$m_\sigma (MeV)$ & 675 & - & - & - \\
$\mu$ & - & 0.45 & 0.57 & 1.00 \\  \hline
%$\delta_{OGE} (MeV)$ & 145.6 & 145.6 \\
%$\delta_{OPE} (MeV)$ & 148.1 & 148.1 \\
\hline
\end{tabular}
\end{center}

\section{Comparison between the Chiral quark model and QDCSM}

To compare these two models, the $NN$ scattering and the properties
of deuteron are calculated. In order to make the intermediate-range
attraction mechanism stand out, one exact the same set (set 1) of
parameters: $b,\alpha_s,\alpha_{ch},m_u,m_\pi,\Lambda$ were used for
both models. Thus, the two models have exactly the same
contributions from one-gluon-exchange and $\pi$ exchange. The only
difference of these two models is how to get the intermediate-range
attraction, $\sigma$ exchange for chiral quark model, quark
delocalization color screening for QDCSM. To test the sensitivity of
the QDCSM to the parameters, other two sets of parameters are used
and the gluon contribution in these two cases is a little different
from the Salamanca ones.

The resonating-group method (RGM) is used in the calculations. The
detail of the method can be found in ref.\cite{RGM}. For QDCSM, we
first fixed baryon size parameter $b$, then the quark-gluon coupling
constant $\alpha_s$ is adjusted to fit the N-$\Delta$ mass
difference, the color screening parameter $\mu$ is then fixed by
deuteron properties. The calculated results for deuteron properties
and $NN$ scattering phase shifts are shown in table II and figs.1-4.

\begin{center}
Table II. The properties of deuteron
\begin{tabular}{cccccc}\hline\hline
&  & Salamanca Model & \multicolumn{3}{c}{QDCSM}  \\
& & & set 1 & set 2 & set 3 \\ \hline
& B (MeV) & 2.0 & 1.94 & 2.01 & 2.01 \\
$d$ & $\sqrt{r^2} (fm)$ & 1.96 & 1.93 & 1.92 & 1.94 \\
 & $P_D (\%)$ & 4.86 & 5.25 & 5.25 & 5.25 \\ \hline
%$d^*$ & B (MeV) & 38 & 198 & 198 & 198 \\
%& $\sqrt{r^2} (fm)$ & 0.97 & 1.0 & 1.0 & 1.0 \\ \hline\hline
\end{tabular}
\end{center}

{\noindent {\em deuteron}: Two models both give a good description
of deuteron. For QDCSM, by adjusting the color screening parameter,
almost the same results for deuteron can be obtained for the three
parameter sets of baryon size $b$. Because of the large separation
between the proton and neutron in the deuteron, the properties of
deuteron mainly reflect the long-range part of the nuclear force.
The same $\pi$-exchange used in the two models assure the properties
of deuteron be fitted equally well. Of course, the enough
intermediate-range attraction is needed to make the deuteron bound.}

\begin{center}
\epsfxsize=3.0in \epsfbox{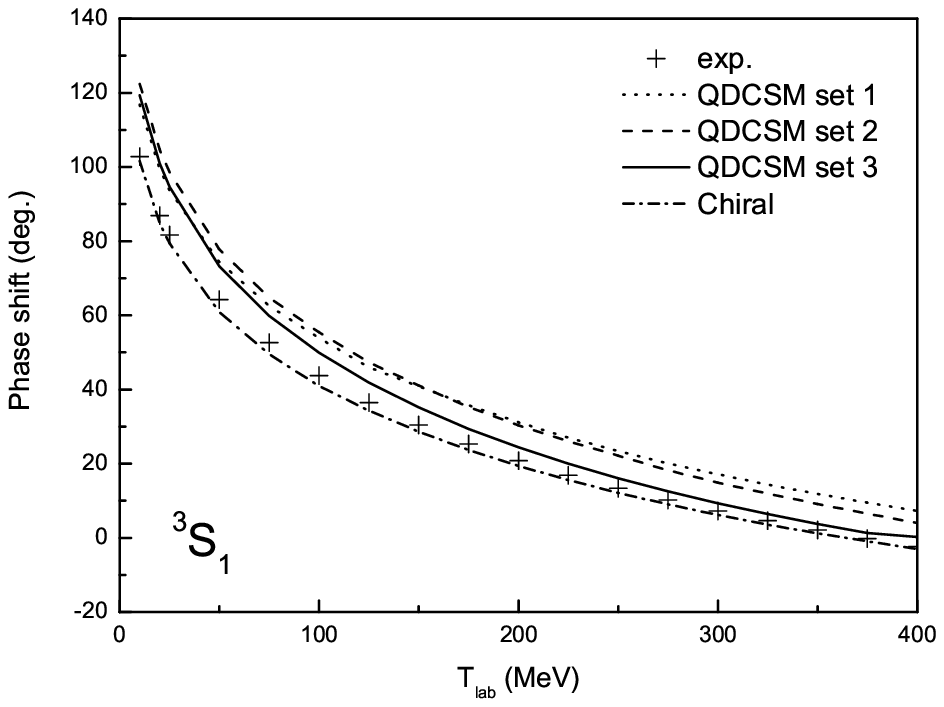}

Fig. 1. The phase shifts for channels $^3S_1$.
\end{center}

\begin{center}
\epsfxsize=3.0in \epsfbox{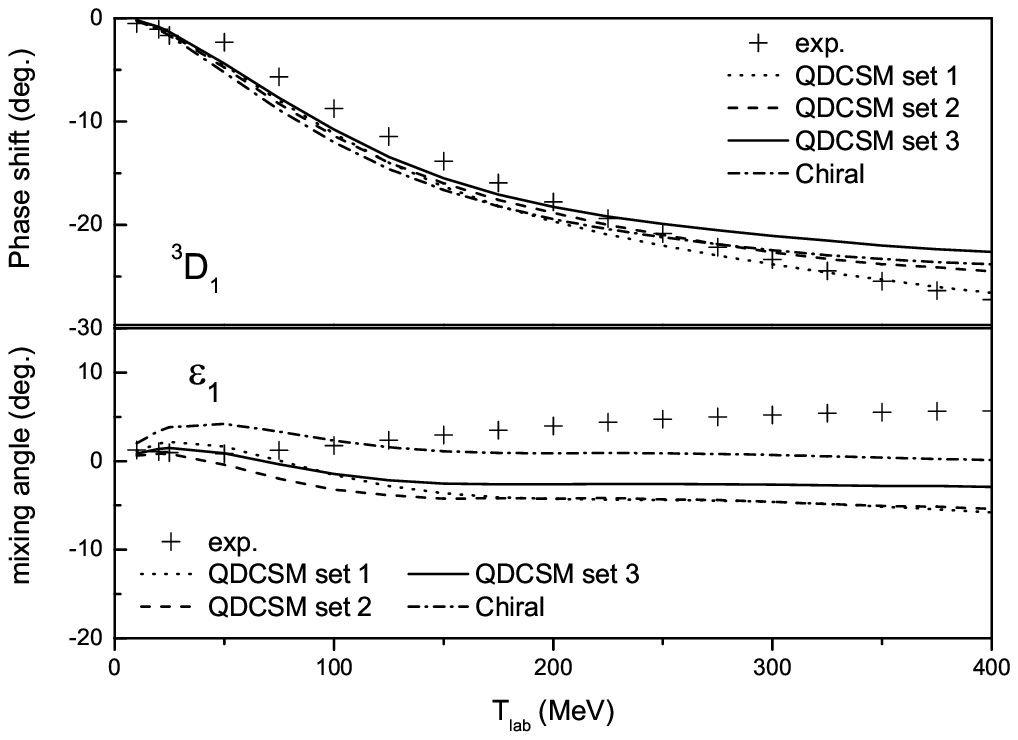}

Fig.2. The phase shifts for channel $^3D_1$ and mixing angles
$\epsilon_1$.
\end{center}

\begin{center}
\epsfxsize=3.0in \epsfbox{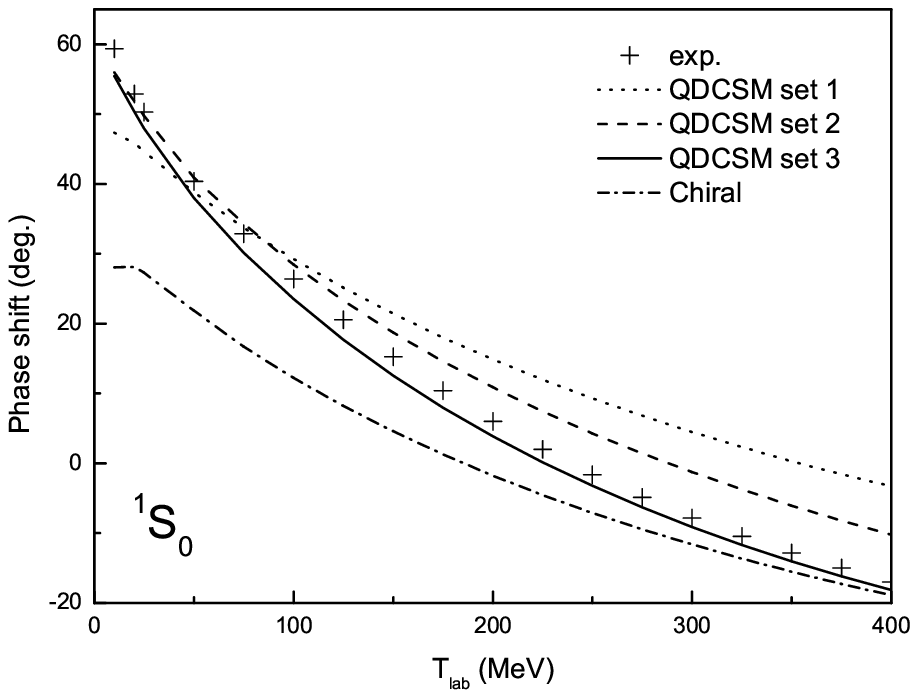}

Fig. 3. The phase shifts for channels $^1S_0$.
\end{center}

\begin{center}
\epsfxsize=3.0in \epsfbox{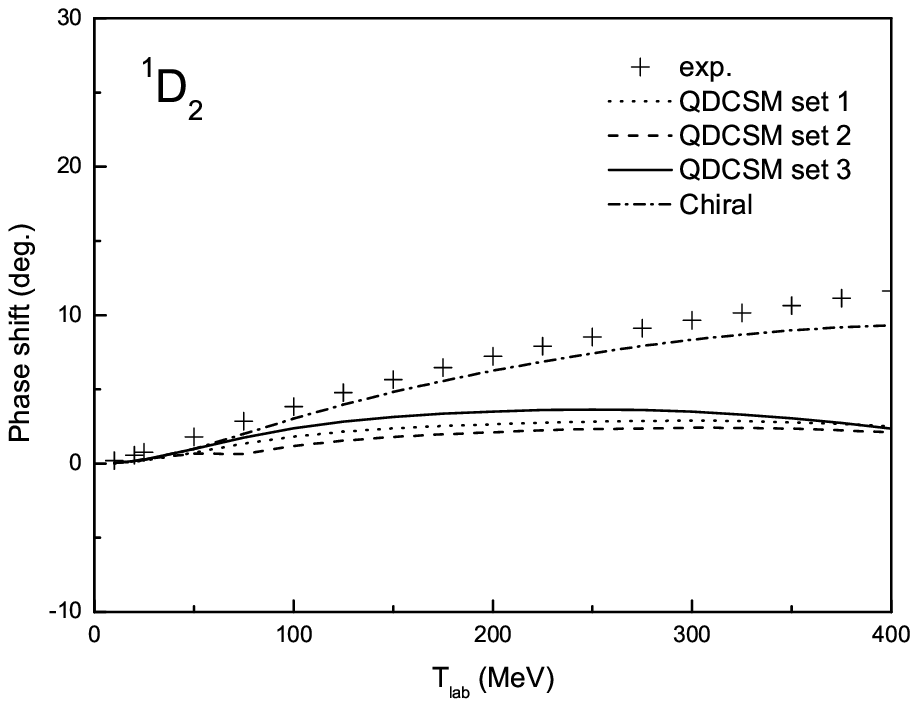}

Fig. 4. The phase shifts for channels $^1D_2$.
\end{center}

{\noindent {\em $NN$ scattering phase shifts: }}\\
$S$-wave: For $^3S_1$, Salamanca model gave an almost perfect
description of the experimental data, QDCSM also had a good
agreement. For $^1S_0$, QDCSM described the experiment data a little
better than Salamanca's single channel approximation did but after
including N$\Delta$ $^1S_0$ channel coupling Salamanca model also
fitted the experimental data well. For QDCSM, the larger the baryon
size $b$, the better the agreement. The dominant contribution to the
$S$-wave phase shift comes from the central part of the potentials,
the agreement between two models means they have the same behavior,
at least for central part of the interaction. \\
$D$-wave: For $^3D_1$, Two models fitted the experimental data
equally. for $^1D_2$, the Salamanca model gave much better fit to
the experimental data than QDCSM did, especially at higher energy.
For QDCSM, different $b$ almost gave same results. \\
Mixing angle $\epsilon_1$: Two models gave a qualitative description
of the experimental data. The chiral quark model result is a little
closer to the experimental ones, but has the same tendency as QDCSM.

\section{Summary}

By calculating the deuteron properties and $NN$ scattering phase
shifts, we compared the QDCSM with the Salamanca version of the
chiral quark model, both models give a good description of the
deuteron and $NN$ scattering, although different intermediate-range
attraction mechanisms were used. The almost same agreement with
experimental data suggest that the $\sigma$-meson exchange can be
replaced by quark delocalization and color screening mechanism. If
one takes the QDCSM mechanism to describe the $NN$ intermediate and
short range interaction, the similarity between molecular and
nuclear force obtained a natural explanation.

\end{document}